\documentclass[sigconf]{acmart}
\AtBeginDocument{%
  }

\copyrightyear{2026}
\acmYear{2026}
\setcopyright{cc}
\setcctype{by}
\acmConference[CIKM '26]{Proceedings of the 35th ACM International Conference on Information and Knowledge Management}{November 07--11, 2026}{Rome, Italy}
\acmBooktitle{Proceedings of the 35th ACM International Conference on Information and Knowledge Management (CIKM '26), November 07--11, 2026, Rome, Italy}
\acmDOI{10.1145/3799682.3839868}
\acmISBN{979-8-4007-2539-5/2026/11}

\usepackage{listings}
\usepackage{xcolor}
\usepackage{framed}
\definecolor{shadecolor}{RGB}{245,245,245}
\usepackage{algorithm}
\usepackage{algpseudocode}
\usepackage{booktabs,subcaption}
\usepackage{adjustbox}
\usepackage{enumitem}

\begin{document}

\title{Ask to Be Sure: Informative Interactions for Confident Multi-Turn LLM Recommendation}

\author{Cedar Site Bai}
\affiliation{%
  \institution{Amazon}
  \city{Sunnyvale}
  \state{CA}
  \country{USA}
}
\email{cedarbai@amazon.com}

\author{Zhenyu Liao}
\affiliation{%
  \institution{Amazon}
  \city{Sunnyvale}
  \state{CA}
  \country{USA}
}
\email{zyliao@amazon.com}

\author{Duanshun Li}
\affiliation{%
  \institution{Amazon}
  \city{Seattle}
  \state{WA}
  \country{USA}
}
\email{duanshun@amazon.com}

\author{Sheikh Sarwar}
\affiliation{%
  \institution{Amazon}
  \city{Sunnyvale}
  \state{CA}
  \country{USA}
}
\email{smsarwar@amazon.com}

\author{Huiyuan Chen}
\affiliation{%
  \institution{Amazon}
  \city{Sunnyvale}
  \state{CA}
  \country{USA}
}
\email{huiyuach@amazon.com}

\author{Yuan Chen}
\affiliation{%
  \institution{Amazon}
  \city{Sunnyvale}
  \state{CA}
  \country{USA}
}
\email{yuanchn@amazon.com}

\author{Changhe Yuan}
\affiliation{%
  \institution{Amazon}
  \city{New York}
  \state{NY}
  \country{USA}
}
\email{ychanghe@amazon.com}

\author{Haiyang Zhang}
\affiliation{%
  \institution{Amazon}
  \city{Sunnyvale}
  \state{CA}
  \country{USA}
}
\email{hhaiz@amazon.com}

\author{Qilin Qi}
\affiliation{%
  \institution{Amazon}
  \city{Sunnyvale}
  \state{CA}
  \country{USA}
}
\email{qilinqi@amazon.com}

\renewcommand{\shortauthors}{Bai et al.}

\begin{abstract}
  Recent advances in large language models (LLMs) have enabled their use as conversational recommender systems (CRS), demonstrating strong recommendation accuracy and natural dialogue. However, guiding multi-turn interactions to elicit user preferences effectively remains challenging. Existing approaches either use separate reinforcement learning agents with templated interactions or optimize for interactivity judged by another LLM, without measuring how much useful information is actually gained. We propose a new approach that quantifies the effectiveness of each interaction by the reduction in the assistant's uncertainty, measured via entropy over recommendations. We apply this entropy reduction as a reward to fine-tune the LLM for strategic interaction generation, without relying on ground-truth recommendations, which are often unavailable in real-world scenarios. Empirical results with supervised fine-tuning (SFT) and direct preference optimization (DPO) on the INSPIRED and ReDial datasets show that our method improves both recommendation quality and conversational efficiency.
\end{abstract}



\maketitle

\section{Introduction}
Conversational recommendation systems (CRS) \citep{sun2018conversational, li2018towards, chen2019towards, zhou2020improving, wang2022towards} have recently garnered increasing attention for their ability to interpret user intent and provide personalized recommendations through natural language interaction. By engaging users in multi-turn dialogues, CRS can elicit preferences, clarify ambiguous requests, and refine recommendations from real-time feedback, transforming recommendation from a one-shot prediction task into an interactive preference discovery process.

With the rapid development of large language models (LLMs), there has been a noticeable shift toward employing LLMs directly as conversational recommenders \citep{he2023cikm, zhu2024collaborative, zhu2025collaborative, zhang2025collm, he2025wsdm}. These systems have shown superior performance over conventional systems \citep{li2018towards, chen2019towards, zhou2020improving} in both recommendation accuracy and user-aware dialogue.

Despite this progress, key challenges remain, including preference elicitation, the integration of user history and collaborative filtering, up-to-date candidate retrieval, and grounding in real-world knowledge. We focus on the first challenge: understanding and eliciting user preferences through strategic, multi-turn interactions, while leaving the others for future exploration. We specifically examine how to design interactions that most effectively solicit user preferences. In an ideal setting, conversational recommendation should be a collaborative process between the user and the AI assistant, rather than a one-way delivery of suggestions.

Existing approaches that use multi-turn dialogue to capture user preferences typically train a separate reinforcement learning (RL) agent to decide when and what to ask or recommend \citep{deng2021unified, du2025sapient}. In such settings, LLMs are used only for dialogue generation, not recommendation, and interaction styles are restricted to rigid formats such as multiple-choice questions or Yes/No responses, limiting natural conversational flow. In parallel, CollabLLM encourages assistant interactivity through LLM-judge rewards \citep{wu2025collabllm}; while effective in making LLMs more proactive in general-purpose tasks, in recommendation contexts it remains unclear how much useful preference information is gained and how these interactions benefit recommendation.

To tackle this challenge, we draw inspiration from recent findings on the correlation between recommendation reliability and uncertainty \citep{kweon2025uncertainty}. That work estimates predictive uncertainty over ranking lists from candidate top-1 probabilities using a Plackett--Luce model and finds that lower uncertainty is associated with better recommendation performance.

We propose a method to quantify the effectiveness of each assistant interaction by the information provided in the corresponding user response. We measure this information gain as the reduction in the assistant’s recommendation uncertainty, computed via entropy over recommendations. This entropy reduction is then used as a reward—without relying on ground-truth recommendations, which are often unavailable in real-world scenarios—to fine-tune the LLM for strategic interaction generation. Empirical results on the INSPIRED and ReDial datasets, using both supervised fine-tuning (SFT) and direct preference optimization (DPO), demonstrate that our method improves both recommendation quality and conversational efficiency.

\noindent\textit{Contributions.}
(1) We introduce an uncertainty-reduction reward that improves conversation strategy and recommendation accuracy.
(2) The metric provides a practical solution when ground-truth recommendations are unavailable.
(3) We contribute two turn-level completion-pair datasets from INSPIRED and ReDial for DPO fine-tuning.

\section{Related Work}
Recent work shows that LLMs remain brittle in multi-turn interaction \citep{laban2025lost}, motivating methods for clarification, preference following, and multi-turn RLHF \citep{chen2025act,zhao2025prefeval,shani2024mtrlpf,zhou2024archer,abdulhai2023lmrlgym,gao2024refuel}. CollabLLM encourages assistants to be proactive through collaboratively generated data and LLM-judge rewards \citep{wu2025collabllm}; however, in recommendation contexts, a response can appear interactive while still revealing little useful preference information. Recommendation-oriented LLM work improves item control, tokenization, retrieval, or collaborative-signal integration \citep{liang2025taxrec,he2025wsdm,zhu2025collaborative,zheng2024lcrec,zhang2025collm}, but these methods mainly improve \emph{what} is recommended given a fixed conversation state rather than how the assistant should change that state through strategic elicitation. Orthogonally, Bayesian teaching studies how models update beliefs under controlled preference-elicitation settings \citep{qiu2025bayesianteaching}. Our focus is targeted and recommendation-specific: we measure \emph{which} assistant turns reduce uncertainty over recommendations, and use that information gain as a reward for open-ended conversational recommendation.

\section{Proposed Method}
In conversational recommendation, the recommender (or AI assistant) engages users in multi-turn dialogues to dynamically elicit preferences, often by asking targeted questions. With each turn, the user provides more information, enabling the assistant to better understand their preferences and become more certain about what to recommend. For example, at the very beginning, the assistant has no prior knowledge of the user and is highly uncertain, resorting to essentially random recommendations. As the dialogue unfolds, the user reveals likes and dislikes, recently watched movies, and other relevant details. To encourage the assistant to ask the most informative questions, we aim for each interaction to significantly reduce recommendation uncertainty. We now introduce our measure of uncertainty.

\subsection{Uncertainty Measurement}
We define uncertainty as the Shannon entropy over the distribution of recommendations induced by repeated sampling of a fixed prompt given a conversation $C$. Following \citet{kweon2025uncertainty}, we use entropy over a recommendation distribution as an uncertainty estimate. Whereas they estimate a distribution over ranking permutations from candidate logits at a fixed conversation state, we empirically estimate an item distribution from repeated top-$m$ list generation, incorporate rank position through weighted entropy, and track how uncertainty changes across conversational interactions. The prompt is as follows:

\begin{shaded}
\small\ttfamily
Given the following conversation history:

\{conversation $C$\}

Generate a list of the top \{$m$\} movies the user would like to watch based on this conversation.
Format your response as a numbered list with no extra sentences:

1. [First movie]
2. [Second movie]
3. [Third movie]

Make sure each recommendation is unique and plausible given the conversation context.
\end{shaded}

\subsubsection{Entropy over recommendations}
We sample $n$ lists $\{L^{(i)}\}_{i=1}^{n}$ generated by the AI assistant based on a conversation $C$, where each list
$L^{(i)} = [\,r^{(i)}_{j}\,]_{j=1}^{m}$ contains the top-$m$ movie recommendations.  
Let $R = \mathrm{Set}\big(\{L^{(i)}\}_{i=1}^{n}\big)$ denote the set of unique recommendations appearing across all lists.
For each item $r \in R$, define its count and empirical frequency as
\begin{align}
c(r) &= \sum_{i=1}^{n}\sum_{j=1}^{m} \mathbf{1}\{\,r^{(i)}_{j} = r\,\}, \\
p(r) &= \frac{c(r)}{m\,n},
\end{align}
which serves as an empirical estimate of the recommendation distribution. Here, $\mathbf{1}\{\cdot\}$ is the indicator function that equals 1 if $r_j^{(i)}=r$ and 0 otherwise.
The entropy over recommendations is calculated as
\begin{equation}
H(C) \;=\; - \sum_{r \in R} p(r)\,\log_{2} p(r).
\end{equation}

\subsubsection{Weighted entropy over recommendations}
To further incorporate the difference in ranking order in each recommendation list $L^{(i)} = [\,r^{(i)}_{j}\,]_{j=1}^{m}$, we apply the following weighted entropy with logarithmic decay. For each rank position $j\in\{1,2,...,m\}$, the weight is
\begin{equation}
w_{j} \;=\; \frac{1}{\log_{2}(j+1)}, \qquad j \in \{1,2,\ldots,m\}.
\end{equation}
For each item $r \in R$, define the weighted count as
\begin{equation}
c_{w}(r) \;=\; \sum_{i=1}^{n}\sum_{j=1}^{m} w_{j}\,\mathbf{1}\{\,r^{(i)}_{j} = r\,\},
\end{equation}
and the corresponding weighted empirical distribution is calculated as
\begin{equation}
p_{w}(r) \;=\; \frac{c_{w}(r)}{\sum_{r' \in R} c_{w}(r')} \;=\; \frac{c_{w}(r)}{\,n \sum_{j=1}^{m} w_{j}\,}.
\end{equation}
The weighted entropy is then
\begin{equation}
H_{w}(C) \;=\; - \sum_{r \in R} p_{w}(r)\,\log_{2} p_{w}(r).
\end{equation}

\subsection{Information Gain as Uncertainty Reduction}
With uncertainty defined, we measure the reduction in uncertainty after an interaction, which we regard as recommendation-relevant information gain. Building on this uncertainty estimator, our contribution is to use its reduction after an assistant turn and the user response it elicits as a reward. This assigns credit to interactions according to recommendation-relevant information gain rather than an LLM judge's general assessment of interactivity \citep{wu2025collabllm}. We introduce two entropy-reduction measures.

\subsubsection{Turn-level entropy reduction}
Turn-level entropy reduction measures the reduction in entropy attributable to one conversational turn. Specifically, given a partial conversation $C_{1} = [U_{1}]$, compute the uncertainty $H_{w}(C_{1})$. An assistant interaction $A_{1}$ is generated (e.g., an initial recommendation or a clarifying question). The user responds with $U_{2}$ (e.g., expressing likes/dislikes or providing preferences). The extended conversation is $C_{2} = [U_{1}, A_{1}, U_{2}]$; compute $H_{w}(C_{2})$. The turn-level information gain is
\begin{equation}
I_{T}(A_{1}) \;=\; H_{w}(C_{1}) \;-\; H_{w}(C_{2}).
\end{equation}

\subsubsection{Conversation-level entropy reduction}
We also evaluate a turn by the total reduction achieved over the full conversation unrolled from that turn. We call this conversation-level entropy reduction. Specifically, given $C_{1} = [U_{1}]$, compute $H_{w}(C_{1})$. Generate an assistant interaction $A_{1}$. Unroll to a terminal conversation $C = [U_{1}, A_{1}, U_{2}, A_{2}, \ldots, A_{\tau}, U_{\tau}]$ and compute $H_{w}(C)$. The conversation-level information gain is
\begin{equation}
I_{C}(A_{1}) \;=\; H_{w}(C_{1}) \;-\; H_{w}(C).
\end{equation}

\subsection{An Illustrative Example}
We now give an example of the trend in entropy reduction in conversational recommendation. We use an INSPIRED conversation written by human workers \citep{hayati2020inspired} and provided in the Supplement \citep{asktobesure2026supplement}. After each user response, we sample $n=5$ lists of top-$m=5$ movie recommendations and calculate the corresponding entropy. As shown in Figure \ref{fig:entropy}, entropy exhibits a decreasing trend as the conversation progresses and the assistant gathers more information from the user.

To verify that this pattern extends beyond one example, we repeat the same experiment over 300 conversations from INSPIRED and average the recommendation entropy at each turn. Figure~\ref{fig:entropy-300} shows that mean entropy generally decreases as conversations progress, with a weighted trend of $-0.040$ bits per turn.

\noindent\begin{minipage}{\columnwidth}
  \centering
  \includegraphics[width=\columnwidth]{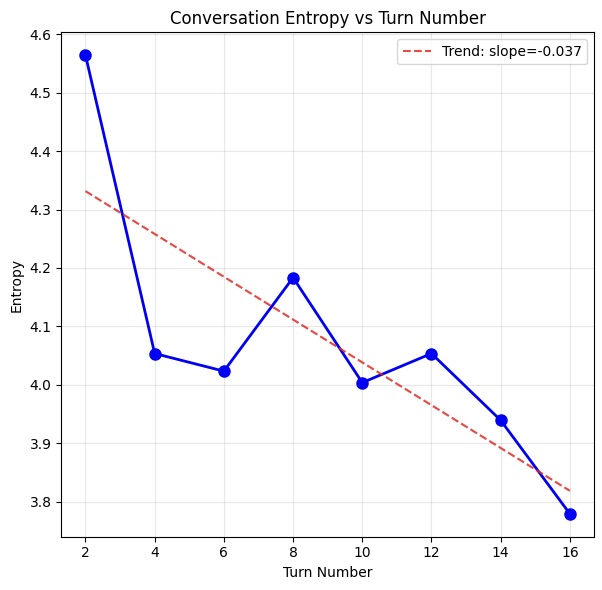}
  \captionsetup{type=figure,hypcap=false}
  \caption{Entropy trend for an INSPIRED conversation; see Supplement \citep{asktobesure2026supplement}.}
  \label{fig:entropy}
  \Description{Line chart showing recommendation entropy generally decreasing as the conversation progresses.}
\end{minipage}

\noindent\begin{minipage}{\columnwidth}
  \centering
  \includegraphics[width=\columnwidth]{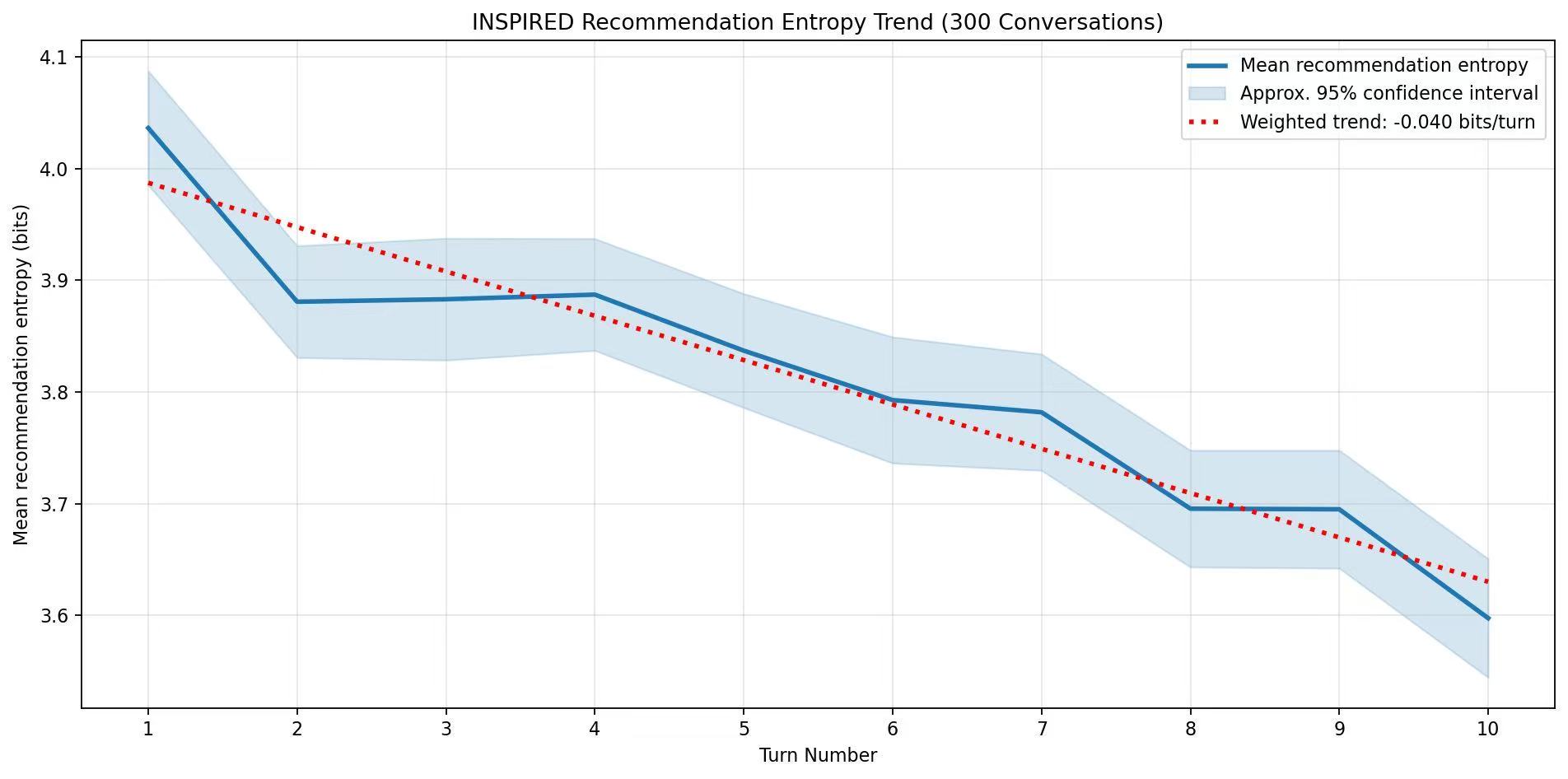}
  \captionsetup{type=figure,hypcap=false}
  \caption{Mean recommendation entropy by turn over 300 INSPIRED conversations. The shaded region shows the approximate 95\% confidence interval.}
  \label{fig:entropy-300}
  \Description{Line chart showing mean recommendation entropy across 300 INSPIRED conversations. Entropy generally decreases from the first through the tenth turn, with a weighted trend of negative 0.040 bits per turn; a shaded band shows the approximate 95 percent confidence interval.}
\end{minipage}

\subsection{Fine-tuning with Entropy Reduction}
We fine-tune LLMs using the proposed entropy-reduction score as the reward. We use standard SFT and DPO \citep{rafailov2023dpo} as optimization mechanisms; the proposed component is the entropy-reduction reward used to select SFT examples or define DPO preferences. For SFT, we select training examples with high reward scores. For DPO, among two generations sampled from the recommender, the higher-reward one is preferred.

\section{Experiments}
\subsection{Baselines}
We begin by introducing the baselines used in the experiments.
We do not directly compare with the RL-based methods in \citep{deng2021unified,du2025sapient}: they learn separate dialogue policies over restricted multiple-choice or Yes/No action spaces, whereas our setting uses a single LLM for open-ended dialogue and recommendation, making their interfaces and reported metrics not directly comparable.
\par\noindent\textbf{Vanilla.} We compare with the vanilla zero-shot recommender \citep{he2023cikm}, i.e., directly using the pre-trained LLM as the recommender.
\par\noindent\textbf{SFT.} We also compare with an LLM fine-tuned via supervised fine-tuning (SFT) on the raw training data from the INSPIRED and ReDial datasets.
\par\noindent\textbf{CollabLLM.} For each fine-tuning approach, we additionally adopt the CollabLLM framework \citep{wu2025collabllm} as a baseline. Its reward combines \textit{task-specific reward} ($R_{\text{task}}$; ground-truth hit for movie recommendation), \textit{interactivity} ($R_{\text{interact}}$; LLM-judge score using the prompt in the Supplement \citep{asktobesure2026supplement}), and \textit{token efficiency} ($R_{\text{tok}}$; generated tokens divided by the maximum allowed token budget):
\begin{equation}
  R \;=\; R_{\text{task}} \;+\; R_{\text{interact}} \;-\; 0.1 \,\cdot\, R_{\text{tok}},
\end{equation}
where \(R_{\text{tok}} = \frac{\#\text{tokens generated}}{\text{max tokens allowed}}\).

\begin{table*}[!t]
\centering
\caption{Results on the INSPIRED and ReDial datasets (best per column in \textbf{bold}).}
\label{tab:res}
\setlength{\tabcolsep}{4pt}
\renewcommand{\arraystretch}{1.05}

\begin{adjustbox}{max width=\textwidth}
\begin{tabular}{lcccccccc}
\toprule
 & \multicolumn{4}{c}{\textbf{INSPIRED Dataset}} & \multicolumn{4}{c}{\textbf{ReDial Dataset}} \\
\cmidrule(lr){2-5} \cmidrule(lr){6-9}
\textbf{Method}
& \textbf{Hit@1} $\uparrow$
& \textbf{Hit@5} $\uparrow$
& \textbf{\shortstack{Simulated\\Conversation Hit}} $\uparrow$
& \textbf{\shortstack{\# Turns to\\Ground Truth}} $\downarrow$
& \textbf{Hit@1} $\uparrow$
& \textbf{Hit@5} $\uparrow$
& \textbf{\shortstack{Simulated\\Conversation Hit}} $\uparrow$
& \textbf{\shortstack{\# Turns to\\Ground Truth}} $\downarrow$ \\
\midrule
Vanilla
& \(1.60\pm0.21\%\) & \(1.94\pm0.18\%\) & \(21.54\pm1.26\%\) & 3.37
& \(0.78\pm0.08\%\) & \(1.73\pm0.15\%\) & \(25.35\pm0.38\%\) & 2.94 \\

SFT (Raw)
& \(2.06\pm0.22\%\) & \(3.32\pm0.14\%\) & \(23.90\pm0.48\%\) & 4.12
& \(1.56\pm0.19\%\) & \(2.96\pm0.26\%\) & \(23.98\pm0.24\%\) & 2.87 \\

SFT (CollabLLM)
& \(2.68\pm0.22\%\) & \(4.26\pm0.14\%\) & \(25.25\pm1.65\%\) & 3.36
& \(2.14\pm0.04\%\) & \(5.00\pm0.09\%\) & \(28.70\pm0.68\%\) & 2.85 \\

SFT (Turn Entropy)
& \(3.00\pm0.22\%\) & \(5.05\pm0.47\%\) & \(26.60\pm0.48\%\) & 3.12
& \(2.17\pm0.06\%\) & \(5.05\pm0.08\%\) & \(28.94\pm0.30\%\) & 2.80 \\

SFT (Conv Entropy)
& \(3.00\pm0.26\%\) & \textbf{\(5.21\pm0.39\%\)} & \(26.93\pm0.48\%\) & \textbf{2.94}
& \(2.16\pm0.05\%\) & \(5.19\pm0.07\%\) & \(29.31\pm0.49\%\) & 2.81 \\

DPO (CollabLLM)
& \(3.15\pm0.34\%\) & \(5.09\pm0.10\%\) & \(26.60\pm1.20\%\) & 3.12
& \(2.22\pm0.03\%\) & \(5.12\pm0.04\%\) & \(30.03\pm0.35\%\) & 2.86 \\

DPO (Turn Entropy)
& \(\mathbf{3.32\pm0.12\%}\) & \(\mathbf{5.21\pm0.22\%}\) & \(\mathbf{27.94\pm0.73\%}\) & 3.07
& \(\mathbf{2.24\pm0.01\%}\) & \(\mathbf{5.21\pm0.04\%}\) & \(31.62\pm0.36\%\) & 2.75 \\

DPO (Conv Entropy)
& \(3.15\pm0.12\%\) & \(5.05\pm0.13\%\) & \(27.02\pm0.55\%\) & 3.05
& \(2.18\pm0.03\%\) & \(5.14\pm0.02\%\) & \(\mathbf{32.83\pm0.31\%}\) & \textbf{2.74} \\
\bottomrule
\end{tabular}
\end{adjustbox}
\end{table*}

\subsection{Experimental Setup}
\noindent\textbf{Datasets.} We use the INSPIRED \citep{hayati2020inspired} and ReDial \citep{li2018towards} conversational recommendation datasets, both collected by human workers on Amazon Mechanical Turk. Besides direct supervised fine-tuning, we use their conversations as references for user simulation. INSPIRED contains 801 training and 99 test conversations, each labeled with a ground-truth recommendation that the user accepts. In ReDial, movies are labeled by whether the user likes and has seen them. We use liked but unseen movies as ground truth and filter examples with no ground truth, yielding 8631 training and 1036 test conversations.

\par\noindent\textbf{Simulation.} We adapt CollabLLM's data generation approach \citep{wu2025collabllm}, in which a user simulator generates conversations with the AI assistant. In this paper, we generate simulated conversations based on INSPIRED and ReDial: Claude Sonnet 4 is prompted with a reference conversation and asked to role-play the user; the partially adapted prompt is in the Supplement \citep{asktobesure2026supplement}. The user terminates when (a) the ground-truth movie is recommended, (b) a satisfactory answer is obtained, (c) the assistant is no longer helpful, or (d) a maximum of 5 turns is reached. We generate datasets and test all methods using Llama-3.2-1B-Instruct \citep{grattafiori2024llama3}.

\par\noindent\textbf{Turn-level DPO pairs.} Following CollabLLM \citep{wu2025collabllm}, we generate turn-level completion pairs for DPO with Algorithm \ref{alg:data_gen}.

\begin{algorithm}[t]
\caption{Simulated Pairwise Selection for Conversation Generation}
\label{alg:data_gen}
\begin{algorithmic}[1]
\Require User simulator $U$; AI assistant $A$; conversation generator $\mathcal{G}$ (based on $U$ and $A$); reward evaluator $R$; reference conversation $C_{\mathrm{ref}}$ (INSPIRED/ReDial); number of turns $T$
\Ensure Generated conversation $C_{\mathrm{gen}}$
\State $C_{\mathrm{gen}} \gets [\,]$ \Comment{initial empty conversation}
\For{$t = 1$ to $T$}
  \State $U_t \gets U(C_{\mathrm{gen}}, C_{\mathrm{ref}})$ \Comment{user simulator emits a query}
  \State $C_{\mathrm{gen}} \gets C_{\mathrm{gen}} \,\|\, [U_t]$ \Comment{append}
  \State $A_t^{(1)} \gets A(C_{\mathrm{gen}})$ \Comment{sample two assistant responses}
  \State $A_t^{(2)} \gets A(C_{\mathrm{gen}})$
  \State $C_t^{(1)} \gets \mathcal{G}\!\big(C_{\mathrm{gen}} \,\|\, [A_t^{(1)}]\big)$ \Comment{full rollout}
  \State $C_t^{(2)} \gets \mathcal{G}\!\big(C_{\mathrm{gen}} \,\|\, [A_t^{(2)}]\big)$
  \State $R_t^{(1)} \gets R(C_t^{(1)})$;\quad $R_t^{(2)} \gets R(C_t^{(2)})$ \Comment{evaluation}
  \If{$R_t^{(1)} \ge R_t^{(2)}$}
    \State \textbf{choose} $A_t^{(1)}$ \textbf{and reject} $A_t^{(2)}$ \Comment{for prompt $C_{\mathrm{gen}}$}
    \State $C_{\mathrm{gen}} \gets C_{\mathrm{gen}} \,\|\, [A_t^{(1)}]$
  \Else
    \State \textbf{choose} $A_t^{(2)}$ \textbf{and reject} $A_t^{(1)}$
    \State $C_{\mathrm{gen}} \gets C_{\mathrm{gen}} \,\|\, [A_t^{(2)}]$
  \EndIf
\EndFor
\end{algorithmic}
\end{algorithm}

Dataset generation uses the CollabLLM reward. The entropy-reduction datasets are generated by relabeling the resulting preference pairs with our entropy-reduction reward.

\par\noindent\textbf{SFT datasets.} SFT (Raw) uses the raw conversation data provided in the training set of INSPIRED (801 conversations) and ReDial (8631 conversations). SFT (CollabLLM) selects conversations with high CollabLLM reward scores from the generated DPO dataset. SFT (Turn Entropy) selects conversations with high entropy-reduction reward scores from the generated DPO dataset. The entropy reduction is measured by the difference between the entropy values before and after the current assistant response and the next user response it elicits. SFT (Conv Entropy) also filters by entropy reduction, measured by the difference between the entropy values before the assistant response and at the end of the whole generated conversation. For fair comparison, we keep the top 801 INSPIRED and 8631 ReDial conversations for each filtered SFT variant.

Across reward variants, we use the same generated preference pairs for relabeling and match the SFT data budget. Thus, within each training paradigm, the comparisons isolate the effect of the reward used for data selection or preference labeling.

\par\noindent\textbf{Evaluation.} For all metrics and baselines, training uses either the INSPIRED/ReDial training set or the generated training set, and testing uses the corresponding test set. Hit@1/Hit@5 follow the tests from \citet{he2023cikm}: each conversation is cut before a ground-truth recommendation, and the recommender is asked to generate a list of 5 movies given the partial conversation; we then calculate and report Hit@1 and Hit@5. A total of 228 INSPIRED and 3552 ReDial partial conversations are evaluated. We also report simulated conversation hit rate by passing each test conversation to a user simulator and checking whether the ground-truth movie is recommended, along with the average number of turns needed to reach that recommendation. This number-of-turns metric reflects whether the recommender generates strategic interactions and efficiently solicits enough information for an accurate recommendation.

\subsection{Example of Baseline and Proposed Reward}
We provide the full generated training example in the Supplement \citep{asktobesure2026supplement}. For the same initial comedy-movie query, two assistant responses received similar CollabLLM-style interactivity scores (0.9 vs. 0.8), but turn-entropy reduction revealed a larger gap (1.728 vs. 0.733). While one might intuitively assume that general questions such as asking about slapstick, witty one-liners, or heartfelt comedy would reveal more about a user's preferences, giving initial recommendations and asking for the user's opinion can also be effective. Most importantly, we do not impose a predefined, subjective notion of what types of questions are good for information gain; our reward leaves this to the observed entropy reduction rather than a manually defined question style.

\subsection{Results}
Table \ref{tab:res} reports the experimental results for each evaluation metric. Models are fine-tuned in the LoRA setting. Throughout testing, the temperature for the assistant is set to 0.1, and we report the average over three runs for each setting.

The entropy-reduction variants show the strongest gains on the main conversational metrics: DPO (Turn Entropy) obtains the best INSPIRED Hit@1, Hit@5, and simulated conversation hit rate, while DPO (Conv Entropy) obtains the best ReDial simulated conversation hit rate and the lowest ReDial turn count.

From Table \ref{tab:res}, we observe that
\begin{itemize}[leftmargin=1.9em,labelwidth=1.45em,labelsep=0.35em,itemindent=0pt,topsep=0pt,itemsep=0pt,parsep=0pt,partopsep=0pt]
    \item [(1)] Direct SFT on the INSPIRED and ReDial datasets can improve hit rate when the assistant is asked to directly generate recommendation lists rather than engage in conversational recommendation, but it does not substantially enhance conversational recommendation ability. This is likely because these datasets are created by human annotators who are not professional recommenders, and their language is often less engaging or proactive than that generated by LLMs.
    \item [(2)] Fine-tuning on simulated data selected via entropy reduction also improves recommendation accuracy, consistent with the finding in \citep{kweon2025uncertainty} that lower recommendation uncertainty is associated with better performance.
    \item [(3)] In the context of conversational recommendation (rather than recommendation list generation) evaluated by simulated conversation hit rate, our proposed entropy-reduction reward outperforms CollabLLM’s reward design—which combines ground-truth hit rate and LLM-judged interactivity—even without access to ground-truth recommendations.
    \item [(4)] Models fine-tuned with the entropy-reduction reward achieve higher conversational efficiency, requiring fewer turns to recommend the ground truth—reflecting effective strategic interaction generation and information gathering.
\end{itemize}
\vskip -0.45\baselineskip
\subsection{Example Conversations Before and After Fine-Tuning}
\vskip -0.15\baselineskip
We provide example conversations before and after our proposed fine-tuning in the Supplement \citep{asktobesure2026supplement}. From these examples, we observe that the fine-tuned model tends to go beyond simply giving recommendations and instead actively asks for user preferences based on the conversation context, thereby improving conversational efficiency.
\subsection{Limitations}
Our evaluation uses automatic recommendation metrics and an LLM-based user simulator rather than human judgments, and we fix $m=n=5$; conversational usefulness and sensitivity to these sampling parameters remain to be validated. We also assume that the base LLM can name candidate items. Dynamic or specialized catalogs would require retrieval, with entropy measured over retrieved rankings.

\section{Conclusion and Future Work}
\vskip -0.15\baselineskip
We introduced an uncertainty-driven approach for multi-turn conversational recommendation that quantifies the assistant's uncertainty as entropy over sampled recommendation lists and rewards turns that reduce this entropy. This objective aligns interaction design with information gain, avoids reliance on ground-truth targets, and integrates cleanly with SFT and DPO. Empirically, on INSPIRED and ReDial, our fine-tuned models improve recommendation quality and conversational efficiency compared to strong baselines. Future work includes evaluating larger models and datasets, integrating the reward into on-policy RLHF methods such as PPO and GRPO, and combining it with CollabLLM's reward when ground-truth recommendations are available. The framework can also be extended to retrieval-based conversational recommendation by measuring entropy over rankings within a retrieved recommendation list \citep{kweon2025uncertainty}.
\label{end:main-content}

\bibliographystyle{ACM-Reference-Format}
\section*{GenAI Usage Disclosure}
\label{start:genai}
The authors used generative AI tools for assistance with language polishing and LaTeX drafting. The authors remain responsible for all content, experiments, claims, and citations in this manuscript.

\bibliography{references}

@String{Computing = "Computing" }

@String{Chelsea = "Chelsea" }

@article{zhang2025collm,
  title={{CoLLM}: Integrating Collaborative Embeddings Into Large Language Models for Recommendation},
  author={Zhang, Yang and Feng, Fuli and Zhang, Jizhi and Bao, Keqin and Wang, Qifan and He, Xiangnan},
  journal={IEEE Transactions on Knowledge and Data Engineering},
  volume={37},
  number={5},
  pages={2329--2340},
  year={2025},
  publisher={IEEE},
  doi={10.1109/TKDE.2025.3540912},
  url={https://doi.org/10.1109/TKDE.2025.3540912}
}

@inproceedings{
zhu2025collaborative,
title={Collaborative Retrieval for Large Language Model-based Conversational Recommender Systems},
author={Yaochen Zhu and Chao Wan and Harald Steck and Dawen Liang and Yesu Feng and Nathan Kallus and Jundong Li},
booktitle={Proceedings of the ACM on Web Conference 2025},
pages={3323--3334},
year={2025},
publisher={Association for Computing Machinery},
address={New York, NY, USA},
doi={10.1145/3696410.3714908},
url={https://doi.org/10.1145/3696410.3714908}
}

@inproceedings{du2025sapient,
    title = "{SAPIENT}: Mastering Multi-turn Conversational Recommendation with Strategic Planning and {M}onte {C}arlo Tree Search",
    author = "Du, Hanwen  and
      Peng, Bo  and
      Ning, Xia",
    editor = "Chiruzzo, Luis  and
      Ritter, Alan  and
      Wang, Lu",
    booktitle = "Proceedings of the 2025 Conference of the Nations of the Americas Chapter of the Association for Computational Linguistics: Human Language Technologies (Volume 1: Long Papers)",
    month = apr,
    year = "2025",
    address = "Albuquerque, New Mexico",
    publisher = "Association for Computational Linguistics",
    url = "https://aclanthology.org/2025.naacl-long.133/",
    doi = "10.18653/v1/2025.naacl-long.133",
    pages = "2629--2648",
    ISBN = "979-8-89176-189-6"
}

@inproceedings{he2023cikm,
author = {He, Zhankui and Xie, Zhouhang and Jha, Rahul and Steck, Harald and Liang, Dawen and Feng, Yesu and Majumder, Bodhisattwa Prasad and Kallus, Nathan and McAuley, Julian},
title = {Large Language Models as Zero-Shot Conversational Recommenders},
year = {2023},
isbn = {9798400701245},
publisher = {Association for Computing Machinery},
address = {New York, NY, USA},
url = {https://doi.org/10.1145/3583780.3614949},
doi = {10.1145/3583780.3614949},
booktitle = {Proceedings of the 32nd ACM International Conference on Information and Knowledge Management},
pages = {720--730},
numpages = {11},
location = {Birmingham, United Kingdom},
series = {CIKM '23}
}

@inproceedings{he2025wsdm,
author = {He, Zhankui and Xie, Zhouhang and Steck, Harald and Liang, Dawen and Jha, Rahul and Kallus, Nathan and McAuley, Julian},
title = {Reindex-Then-Adapt: Improving Large Language Models for Conversational Recommendation},
year = {2025},
isbn = {9798400713293},
publisher = {Association for Computing Machinery},
address = {New York, NY, USA},
url = {https://doi.org/10.1145/3701551.3703573},
doi = {10.1145/3701551.3703573},
booktitle = {Proceedings of the Eighteenth ACM International Conference on Web Search and Data Mining},
pages = {866--875},
numpages = {10},
location = {Hannover, Germany},
series = {WSDM '25}
}

@inproceedings{sun2018conversational,
author = {Sun, Yueming and Zhang, Yi},
title = {Conversational Recommender System},
year = {2018},
isbn = {9781450356572},
publisher = {Association for Computing Machinery},
address = {New York, NY, USA},
url = {https://doi.org/10.1145/3209978.3210002},
doi = {10.1145/3209978.3210002},
booktitle = {The 41st International ACM SIGIR Conference on Research \& Development in Information Retrieval},
pages = {235--244},
numpages = {10}
}

@inproceedings{li2018towards,
  title={Towards Deep Conversational Recommendations},
  author={Li, Raymond and Ebrahimi Kahou, Samira and Schulz, Hannes and Michalski, Vincent and Charlin, Laurent and Pal, Chris},
  booktitle={Advances in Neural Information Processing Systems},
  volume={31},
  year={2018},
  publisher={Curran Associates, Inc.},
  url={https://proceedings.neurips.cc/paper/2018/hash/800de15c79c8d840f4e78d3af937d4d4-Abstract.html}
}

@inproceedings{wang2022towards,
author = {Wang, Xiaolei and Zhou, Kun and Wen, Ji-Rong and Zhao, Wayne Xin},
title = {Towards Unified Conversational Recommender Systems via Knowledge-Enhanced Prompt Learning},
year = {2022},
isbn = {9781450393850},
publisher = {Association for Computing Machinery},
address = {New York, NY, USA},
url = {https://doi.org/10.1145/3534678.3539382},
doi = {10.1145/3534678.3539382},
booktitle = {Proceedings of the 28th ACM SIGKDD Conference on Knowledge Discovery and Data Mining},
pages = {1929--1937},
numpages = {9}
}

@inproceedings{zhou2020improving,
  title={Improving Conversational Recommender Systems via Knowledge Graph based Semantic Fusion},
  author={Zhou, Kun and Zhao, Wayne Xin and Bian, Shuqing and Zhou, Yuanhang and Wen, Ji-Rong and Yu, Jingsong},
  booktitle={Proceedings of the 26th ACM SIGKDD International Conference on Knowledge Discovery \& Data Mining},
  pages={1006--1014},
  year={2020},
  publisher={Association for Computing Machinery},
  address={New York, NY, USA},
  doi={10.1145/3394486.3403143},
  url={https://doi.org/10.1145/3394486.3403143}
}

@inproceedings{chen2019towards,
  title={Towards Knowledge-Based Recommender Dialog System},
  author={Chen, Qibin and Lin, Junyang and Zhang, Yichang and Ding, Ming and Cen, Yukuo and Yang, Hongxia and Tang, Jie},
  booktitle={Proceedings of the 2019 Conference on Empirical Methods in Natural Language Processing and the 9th International Joint Conference on Natural Language Processing (EMNLP-IJCNLP)},
  pages={1803--1813},
  year={2019},
  address={Hong Kong, China},
  publisher={Association for Computational Linguistics},
  doi={10.18653/v1/D19-1189},
  url={https://aclanthology.org/D19-1189/}
}

@inproceedings{deng2021unified,
  title={Unified conversational recommendation policy learning via graph-based reinforcement learning},
  author={Deng, Yang and Li, Yaliang and Sun, Fei and Ding, Bolin and Lam, Wai},
  booktitle={Proceedings of the 44th International ACM SIGIR Conference on Research and Development in Information Retrieval},
  pages={1431--1441},
  year={2021},
  publisher={Association for Computing Machinery},
  address={New York, NY, USA},
  doi={10.1145/3404835.3462913},
  url={https://doi.org/10.1145/3404835.3462913}
}

@inproceedings{
zhu2024collaborative,
title={Collaborative Large Language Model for Recommender Systems},
author={Yaochen Zhu and Liang Wu and Qi Guo and Liangjie Hong and Jundong Li},
booktitle={Proceedings of the ACM Web Conference 2024},
pages={3162--3172},
year={2024},
publisher={Association for Computing Machinery},
address={New York, NY, USA},
doi={10.1145/3589334.3645347},
url={https://doi.org/10.1145/3589334.3645347}
}

@inproceedings{hayati2020inspired,
  title={{INSPIRED}: Toward Sociable Recommendation Dialog Systems},
  author={Hayati, Shirley Anugrah and Kang, Dongyeop and Zhu, Qingxiaoyang and Shi, Weiyan and Yu, Zhou},
  booktitle={Proceedings of the 2020 Conference on Empirical Methods in Natural Language Processing (EMNLP)},
  pages={8142--8152},
  year={2020},
  address={Online},
  publisher={Association for Computational Linguistics},
  doi={10.18653/v1/2020.emnlp-main.654},
  url={https://aclanthology.org/2020.emnlp-main.654/}
}

@inproceedings{wu2025collabllm,
  title = {{C}ollab{LLM}: From Passive Responders to Active Collaborators},
  author = {Wu, Shirley and Galley, Michel and Peng, Baolin and Cheng, Hao and Li, Gavin and Dou, Yao and Cai, Weixin and Zou, James and Leskovec, Jure and Gao, Jianfeng},
  booktitle = {Proceedings of the 42nd International Conference on Machine Learning},
  pages = {67260--67283},
  year = {2025},
  volume = {267},
  series = {Proceedings of Machine Learning Research},
  publisher = {PMLR},
  url = {https://proceedings.mlr.press/v267/wu25i.html}
}

@inproceedings{
kweon2025uncertainty,
title={Uncertainty Quantification and Decomposition for {LLM}-based Recommendation},
author={Wonbin Kweon and Sanghwan Jang and SeongKu Kang and Hwanjo Yu},
booktitle={Proceedings of the ACM on Web Conference 2025},
pages={4889--4901},
year={2025},
publisher={Association for Computing Machinery},
address={New York, NY, USA},
doi={10.1145/3696410.3714601},
url={https://doi.org/10.1145/3696410.3714601}
}

@article{grattafiori2024llama3,
  title   = {The Llama 3 Herd of Models},
  author  = {Grattafiori, Aaron and others},
  journal = {arXiv preprint arXiv:2407.21783},
  year    = {2024},
  url     = {https://arxiv.org/abs/2407.21783}
}

@inproceedings{rafailov2023dpo,
  title   = {Direct Preference Optimization: Your Language Model is Secretly a Reward Model},
  author  = {Rafailov, Rafael and Sharma, Archit and Mitchell, Eric and Manning, Christopher D. and Ermon, Stefano and Finn, Chelsea},
  booktitle = {Advances in Neural Information Processing Systems},
  volume  = {36},
  pages   = {53728--53741},
  year    = {2023},
  doi     = {10.52202/075280-2338},
  url     = {https://proceedings.neurips.cc/paper_files/paper/2023/hash/a85b405ed65c6477a4fe8302b5e06ce7-Abstract-Conference.html}
}

@inproceedings{laban2025lost,
  title   = {LLMs Get Lost In Multi-Turn Conversation},
  author  = {Laban, Philippe and Hayashi, Hiroaki and Zhou, Yingbo and Neville, Jennifer},
  booktitle = {The Fourteenth International Conference on Learning Representations},
  year    = {2026},
  url     = {https://openreview.net/forum?id=VKGTGGcwl6}
}

@inproceedings{chen2025act,
  title     = {Learning to Clarify: Multi-turn Conversations with Action-Based Contrastive Self-Training},
  author    = {Chen, Maximillian and Sun, Ruoxi and Pfister, Tomas and Ar{\i}k, Sercan {\"O}.},
  booktitle = {The Thirteenth International Conference on Learning Representations},
  year      = {2025},
  url       = {https://openreview.net/forum?id=SIE6VFps9x}
}

@inproceedings{zhao2025prefeval,
  title     = {Do LLMs Recognize Your Preferences? Evaluating Personalized Preference Following in LLMs},
  author    = {Zhao, Siyan and Hong, Mingyi and Liu, Yang and Hazarika, Devamanyu and Lin, Kaixiang},
  booktitle = {The Thirteenth International Conference on Learning Representations},
  year      = {2025},
  url       = {https://openreview.net/forum?id=QWunLKbBGF}
}

@inproceedings{shani2024mtrlpf,
  title   = {Multi-turn Reinforcement Learning with Preference Human Feedback},
  author  = {Shani, Lior and Rosenberg, Aviv and Cassel, Asaf and Lang, Oran and Calandriello, Daniele and Zipori, Avital and Noga, Hila and Keller, Orgad and Piot, Bilal and Szpektor, Idan and Hassidim, Avinatan and Matias, Yossi and Munos, R{\'e}mi},
  booktitle = {Advances in Neural Information Processing Systems},
  volume  = {37},
  pages   = {118953--118993},
  year    = {2024},
  doi     = {10.52202/079017-3779},
  url     = {https://proceedings.neurips.cc/paper_files/paper/2024/hash/d77a7b289361abff82bdd2fb537ae152-Abstract-Conference.html}
}

@inproceedings{zhou2024archer,
  title     = {ArCHer: Training Language Model Agents via Hierarchical Multi-Turn RL},
  author    = {Zhou, Yifei and Zanette, Andrea and Pan, Jiayi and Levine, Sergey and Kumar, Aviral},
  booktitle = {Proceedings of the 41st International Conference on Machine Learning (ICML)},
  series    = {Proceedings of Machine Learning Research},
  volume    = {235},
  pages     = {62178--62209},
  year      = {2024},
  publisher = {PMLR},
  url       = {https://proceedings.mlr.press/v235/zhou24t.html}
}

@article{abdulhai2023lmrlgym,
  title   = {LMRL Gym: Benchmarks for Multi-Turn Reinforcement Learning with Language Models},
  author  = {Abdulhai, Marwa and White, Isadora and Snell, Charlie and Sun, Charles and Hong, Joey and Zhai, Yuexiang and Xu, Kelvin and Levine, Sergey},
  journal = {arXiv preprint arXiv:2311.18232},
  year    = {2023},
  url     = {https://arxiv.org/abs/2311.18232}
}

@inproceedings{gao2024refuel,
  title   = {Regressing the Relative Future: Efficient Policy Optimization for Multi-turn RLHF},
  author  = {Gao, Zhaolin and Zhan, Wenhao and Chang, Jonathan D. and Swamy, Gokul and Brantley, Kiant{\'e} and Lee, Jason D. and Sun, Wen},
  booktitle = {The Thirteenth International Conference on Learning Representations},
  year    = {2025},
  url     = {https://openreview.net/forum?id=cVyELMpMRS}
}

@inproceedings{liang2025taxrec,
  title     = {Taxonomy-Guided Zero-Shot Recommendations with LLMs},
  author    = {Liang, Yueqing and Yang, Liangwei and Wang, Chen and Xu, Xiongxiao and Yu, Philip S. and Shu, Kai},
  booktitle = {Proceedings of the 31st International Conference on Computational Linguistics (COLING)},
  pages     = {1520--1530},
  year      = {2025},
  address   = {Abu Dhabi, UAE},
  publisher = {Association for Computational Linguistics},
  url       = {https://aclanthology.org/2025.coling-main.102/}
}

@inproceedings{zheng2024lcrec,
  title     = {Adapting Large Language Models by Integrating Collaborative Semantics for Recommendation},
  author    = {Zheng, Bowen and Hou, Yupeng and Lu, Hongyu and Chen, Yu and Zhao, Wayne Xin and Chen, Ming and Wen, Ji-Rong},
  booktitle = {Proceedings of the IEEE International Conference on Data Engineering (ICDE)},
  year      = {2024},
  pages     = {1435--1448},
  publisher = {IEEE},
  doi       = {10.1109/ICDE60146.2024.00118},
  url       = {https://doi.org/10.1109/ICDE60146.2024.00118}
}

@article{qiu2025bayesianteaching,
  title   = {Bayesian Teaching Enables Probabilistic Reasoning in Large Language Models},
  author  = {Qiu, Linlu and Sha, Fei and Allen, Kelsey and Kim, Yoon and Linzen, Tal and van Steenkiste, Sjoerd},
  journal = {Nature Communications},
  volume  = {17},
  number  = {1},
  pages   = {1238},
  year    = {2026},
  doi     = {10.1038/s41467-025-67998-6},
  url     = {https://doi.org/10.1038/s41467-025-67998-6}
}

@misc{asktobesure2026supplement,
  title = {Supplement for ``Ask to Be Sure: Informative Interactions for Confident Multi-Turn LLM Recommendation''},
  author = {Bai, Cedar Site and Li, Duanshun and Liao, Zhenyu and Sarwar, Sheikh and Chen, Huiyuan and Chen, Yuan and Yuan, Changhe and Zhang, Haiyang and Qi, Qilin},
  year = {2026},
  url = {https://github.com/best99317/multiturn_rl/blob/master/supplementary-material.pdf}
}

\end{document}